\documentclass[letterpaper,aps,prb,twocolumn,amsmath,showpacs,amssymb]{revtex4}

\usepackage{epsfig}
\usepackage{graphicx}% Include figure files
\usepackage{dcolumn}% Align table columns on decimal point
\usepackage{bm}% bold math
\usepackage{bbm}
\usepackage{wasysym}
\usepackage{marvosym}
\usepackage{times,amsmath,amssymb}

\newcommand{\trG}{\mathbbm G}

\newlength{\textwidthm}

\begin{document}

\title{Conductivity of suspended and non-suspended graphene at finite gate voltage}

\author{T.~Stauber,$^1$ N.~M.~R.~Peres,$^1$ and A. H. Castro Neto$^2$} 

\affiliation{$^1$Centro de F\'{\i}sica  e  Departamento de
F\'{\i}sica, Universidade do Minho, P-4710-057, Braga, Portugal}

\affiliation{$^2$Department of Physics, Boston University, 590 
Commonwealth Avenue, Boston, MA 02215, USA}

\date{\today}

\begin{abstract}
We compute the DC and the optical conductivity of graphene for finite
values of the chemical potential by taking into account the effect of
disorder, due to mid-gap states (unitary scatterers) and charged impurities, and the effect
of both optical and acoustic phonons.  The disorder due to mid-gap
states is treated in the coherent potential approximation (CPA, a self-consistent approach
based on the Dyson equation),
whereas that due to charged impurities is also treated via the Dyson
equation, with the self-energy computed using second order
perturbation theory.  The effect of the phonons is also included via
the Dyson equation, with the self energy computed using first order
perturbation theory. The self-energy due to phonons is computed both
using the bare electronic Green's function and the full electronic
Green's function, although we show that the effect of disorder on the phonon-propagator is
negligible. Our results are in qualitative agreement with recent
experiments. Quantitative agreement could be obtained if one assumes
water molelcules under the graphene substrate. We also comment on the
electron-hole asymmetry observed in the DC conductivity of suspended
graphene.
\end{abstract}


\maketitle

%----------------------------------------------------------------------------%
% Section
%----------------------------------------------------------------------------%
\section{Introduction} 
The isolation of a single carbon layer via micromechanical cleavage
has triggered immense research activity.\cite{pnas} Apart from the
anomalous quantum Hall effect due to chiral Dirac-like
quasi-particles,\cite{Nov05,Kim05} the finite ``universal'' DC
conductivity at the neutrality point attracted major
attention.\cite{Nov04} For recent reviews see
Refs. \onlinecite{Nov07,Katsnelson07,peresworld,GeimPhysToday,rmp,rmpBeenakker}.

The electronic properties of graphene are characterized by two
nonequivalent Fermi-surfaces around the $K$ and $K'$-points,
respectively, which shrink to two points at the neutrality point
$\mu=0$ ($\mu$ is chemical potential). The spectrum around these two
points is given by an (almost) isotropic energy dispersion $E({\bm
k})=\pm v_F\hbar k$ with the Fermi velocity $v_F\simeq10^6$
m/s.\cite{Wallace} Graphene can thus be described by an effective
(2+1)-dimensional relativistic field theory with the velocity of light
$c$ replaced by the Fermi velocity $v_F$.\cite{Semenoff84}

Relativistic field theories in (2+1) dimensions were investigated long
before the actual discovery of graphene\cite{Haldane88,Ludwig94} and
also the two values of the universal conductivities of a clean system
at the neutrality point depending on whether one includes a broadening
$\Gamma\rightarrow0$ or not were reported
then.\cite{Ryu06,Ziegler07,GusyninIJMPB} In the first case, one
obtains
$\sigma_{\Gamma\rightarrow0}^{\mu=0}=\frac{4}{\pi}e^2/h$,\cite{Gorbar02}
the second case yields
$\sigma_{\Gamma=0}^{\mu=0}=\frac{\pi}{2}e^2/h$.\cite{Falkovsky07}
Interestingly, the first value is also obtained without the limit
$\Gamma\rightarrow0$ within the self-consistent coherent potential
approximation (CPA).\cite{PeresPRB} We also note that the constant
conductivity holds for zero temperature, only; for finite temperature
the DC conductivity is zero.\cite{PeresIJMPB}

If leads are attached to the graphene sample, an external broadening is
introduced and the conductivity is given by
$\sigma_{\Gamma\rightarrow0}^{\mu=0}$
\cite{Katsnelson06,Beenakker06,Verges07}
which has been experimentally verified for samples with large aspect
ratio.\cite{Lau07} This is in contrast to measurements of the optical
conductivity, where leads are absent and a finite energy scale
given by the frequency $\omega$ of the incoming beam renders the
intrinsic disorder negligible, $\Gamma/\hbar\omega\approx0$. 
One thus expects the universal conductivity to be given by 
$\sigma_{\Gamma=0}^{\mu=0}=\frac{\pi}{2}e^2/h$, 
which was measured in various experiments in graphene on a
SiO$_2$,\cite{Basov} SiC-substrate\cite{George} and free
hanging.\cite{GeimOptics} Also in graphene  bilayer and 
multilayers,\cite{George,GeimOptics} as well as in
graphite\cite{Kuzmenko} the conductivity per plane is of the
order of $\sigma_0\equiv\frac{\pi}{2} e^2/h$.

The above results were obtained from the Kubo or Laundauer formula and
assumed coherent transport.  Also diffusive models based on the
semi-classical Boltzmann approach yield a finite DC conductivity at
the neutrality point.  Nevertheless, the finite conductivity was found to be
non-universal\cite{Nomura06,Adam07,PeresBZ,StauberBZ,Kim07} in
contraditions to the findings of early experiments, which suggested
$\sigma_{\rm min}\approx 4e^2/h$.\cite{Nov04} 
We should however stress that one can still assume a certain degree
of {\it universality}, since the experimental values for the
conductivity are all of the order of $4e^2/h$.
It was argued that
electron-hole puddles\cite{Cheianov07} or potential fluctuations in
the substrate\cite{Adam07} can account for a finite conductivity at
the Dirac point. An alternative explanation of this quasi-universal
behavior seen in experiments is that there is only a logarithmic
dependence on the impurity concentration due to mid-gap states and
therefore only in cleaner samples deviations from the universal value
are seen.\cite{StauberBZ}

On the other hand, the optical conductivity is given by the universal conductivity
$\sigma_0$ for frequencies larger than twice the chemical potential
$\mu$. It is remarkable that this universal value also holds in the 
optical frequency range,\cite{GeimOptics,Stauber08} 
a result with important consequences in applications.\cite{ap1,ap2,ap3}
Only for frequencies
$\hbar\omega<2\mu$, the sample-dependent scattering behavior of the
electrons becomes important and recent experiments in show an decay of
the universal conductivity with unusual large broadening around $2\mu$
which can not be explained by thermal effects.\cite{Basov} Moreover,
the spectral weight for $k_BT\ll\hbar\omega\ll2\mu$ does not reach
zero as would be expected due to Pauli blocking, but assumes an almost
constant plateau of $\sigma\approx\sigma_0/3$ for larger gate voltage.

The first calculations of the optical conductivity of graphene, using
the Dirac Hamiltonian were done in Ref. [\onlinecite{PeresPRB}]. 
This study was subsequently revisited
a number of times, \cite{Gusynin06PRL,Gusynin07PRL,Gusynin07} and
summarized in Ref.  [\onlinecite{GusyninIJMPB}]. In these calculations
the effect of disorder was treated  in a  phenomenological manner, by
broadening the delta functions into Lorentzians characterized by
constant width $\Gamma$. As shown in Ref. [\onlinecite{PeresPRB}] however,
the momentum states are non-uniformly broadened, with the states
close to the Dirac point being much more affected by the impurities
than those far away from that point.
In the clean limit, the exact calculation of the
optical properties of graphene was considered in Ref. [\onlinecite{Pedersen}],
a calculation recently generalized to the calculation of
the optical properties of graphene antidot lattices.\cite{pedersen2}

In this paper, we generalize the results of Ref.
[\onlinecite{PeresPRB}] by considering a finite chemical potential, including the
effect of charge impurities, and the scattering by phonons.
We discuss two main corrections to the clean system and calculate the
optical conductivity. First, we include the coupling of the Dirac
fermions to in-plane phonons, acoustical as well as optical ones.
Out-of-plane phonons only have a negligible effect on the electronic
properties of graphene.\cite{StauberHolstein08} Secondly, we include
various types of disorder which give rise to mid-gap states as well as
Coulomb scatterers.

In Sec. \ref{hamilt}, we define the phonon Hamiltonian, 
deduce the electron-phonon interaction and calculate the 
electronic self-energy. In Sec. \ref{GC}, we discuss the Green's function
which is modified due to impurities and phonons. 
We then present our results for DC and 
optical conductivity and compare it to the experiment 
of Ref. [\onlinecite{Basov}]. We close with remarks and conclusions.

%----------------------------------------------------------------------------%
% Section
%----------------------------------------------------------------------------%
\section{Electrons and Phonons} 
\label{hamilt}
\subsection{Tight-binding Hamiltonian and current operator}

The Hamiltonian, in tight binding form, for electrons in graphene
is written as
\begin{eqnarray}
H&=&-\sum_{\bm R,\sigma}\sum_{\bm \delta}t(\bm R,\bm R+\bm \delta)
[a^\dag_\sigma(\bm R)b_\sigma(\bm R+\bm \delta)+H.c.]
\label{TB}
\end{eqnarray} 
where the operator $a^\dag_\sigma(\bm R)$ creates an electron in the
carbon atoms of sub-lattice $A$, and $b^\dag_\sigma(\bm R)$
does the same in sub-lattice $B$. The hopping
parameter, $t(\bm R,\bm R+\bm \delta)$, depends on the
relative position of the carbon atoms both due to the presence of a vector
potential $\bm A(t)$ and due to the vibration of the carbon atoms.
The vectors $\bm\delta$ have the form
\begin{equation}
\begin{array}{l}
\displaystyle{\boldsymbol{\delta}_1=\frac{a}{2}\left(1,\sqrt{3}\right) \qquad
\boldsymbol{\delta}_2=\frac{a}{2}\left(1,-\sqrt{3}\right) \qquad 
\boldsymbol{\delta}_3= -a\left(1,0\right)}\,,
\end{array}
\end{equation}
where $a$ is the carbon-carbon distance.
In order to obtain the current operator we write the 
hopping parameter as
\begin{equation}
t\rightarrow te^{i(e/\hbar)\bm A(t)\cdot\bm\delta}\,.
\end{equation}
 Expanding the exponential up to second order
in the vector potential $\bm A(t)$,and assuming that the electric
field is oriented along the $x$ direction, the current operator is obtained
from
\begin{equation}
j_x=-\frac{\partial H}{\partial A_x(t)}\,,
\end{equation}
leading to $j_x=j_x^P+A_x(t)j^D_x$. The operator $j_x^P$ reads
\begin{eqnarray}
j_x^P&=&\frac {tie}{\hbar}
\sum_{\bm R,\sigma}\sum_{\bm \delta}
[\delta_x a^\dag_\sigma(\bm R)b_\sigma(\bm R+\bm \delta)- H.c.]\,.
\end{eqnarray}
The current term of the operator proportional to $A_x(t)$ will not be used in this
paper, and therefore it is pointless to give its form here.\cite{Stauber08,PeresIJMPB}
%----------------------------------------------------------------------------%
% Section
%----------------------------------------------------------------------------%
\subsection{Phonon modes} 

\begin{figure}[!ht]
\begin{center}
\includegraphics*[scale=0.3]{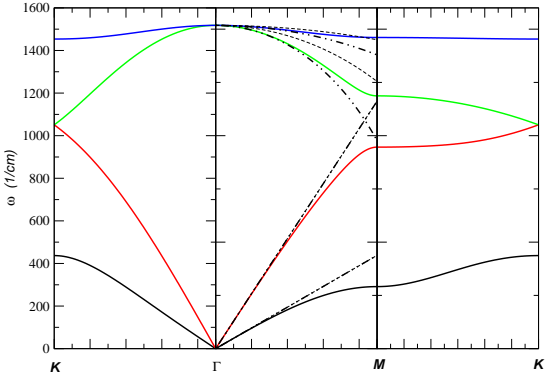}
\end{center}
\caption{
(color online) Phonon spectrum of the honeycomb lattice, using
$\alpha=500$ N/m and $\beta=10$ N/m, such that the optical
frequency at the $\Gamma$-point is or the order
of 1600 cm$^{-1}$. Also shown are the 
spectra of the simplified model discussed in the text (dashed lines)
and the effective model discussed in  Appendix \ref{AA}
(dashed-dotted lines).
\label{phonon}}
\end{figure}

In order to describe the effect of phonons in graphene we adopt the model
developed by Woods and Mahan\cite{Woods} and extensively
used by other authors.\cite{Suzuura,Ando06,Ishikawa,Neto}
The potential energy of the model is a sum of two terms. The first
is due to bond-stretching and reads 
\begin{equation}
V_1=\frac {\alpha} 2 \sum_{\bm R}\sum_{\delta}
[\bm u_A(\bm R)-u_B(\bm R+\bm\delta)\cdot\bm\delta/a]^2\,,
\label{V}
\end{equation}
where $u_A(\bm R)$ and $u_B(\bm R+\bm\delta)$ represent the small
displacements relatively to the equilibrium position of the carbon atoms
in the sub-lattice $A$ and sub-lattice $B$, respectively.
If only the term (\ref{V}) is used a simple analytical expression
for the eigen-modes is obtained.\cite{Cserti} The second term of the model
is due to angle deformation and has the form
\begin{widetext}
\begin{eqnarray}
V_2&=&\frac{\beta}{2a^4}\sum_{\bm R}
\left\{
[(\bm\delta_1/2+\bm\delta_2)\cdot(\bm u_A(\bm R)-\bm u_B(\bm R+\bm\delta_1))
+
(\bm\delta_2/2+\bm\delta_1)\cdot(\bm u_A(\bm R)-\bm u_B(\bm R+\bm\delta_2))
]^2
\right.
\nonumber\\
&+&
[(\bm\delta_1/2+\bm\delta_3)\cdot(\bm u_A(\bm R)-\bm u_B(\bm R+\bm\delta_1))
+
(\bm\delta_3/2+\bm\delta_1)\cdot(\bm u_A(\bm R)-\bm u_B(\bm R+\bm\delta_3))
]^2
\nonumber\\
&+&
[(\bm\delta_2/2+\bm\delta_3)\cdot(\bm u_A(\bm R)-\bm u_B(\bm R+\bm\delta_2))
+
(\bm\delta_3/2+\bm\delta_2)\cdot(\bm u_A(\bm R)-\bm u_B(\bm R+\bm\delta_3))
]^2
\nonumber\\
&+&
[(\bm\delta_1/2+\bm\delta_2)\cdot(\bm u_B(\bm R+\bm\delta_3)-\bm u_A(\bm R-\bm a_1))
+
(\bm\delta_2/2+\bm\delta_1)\cdot(\bm u_B(\bm R+\bm\delta_3)-\bm u_A(\bm R-\bm a_2))
]^2
\nonumber\\
&+&
[(\bm\delta_1/2+\bm\delta_3)\cdot(\bm u_B(\bm R+\bm\delta_3)-\bm u_A(\bm R-\bm a_1))
+
(\bm\delta_3/2+\bm\delta_1)\cdot(\bm u_B(\bm R+\bm\delta_3)-\bm u_A(\bm R))
]^2
\nonumber\\
&+&
\left.
[(\bm\delta_2/2+\bm\delta_3)\cdot(\bm u_B(\bm R+\bm\delta_3)-\bm u_A(\bm R-\bm a_2))
+
(\bm\delta_3/2+\bm\delta_2)\cdot(\bm u_B(\bm R+\bm\delta_3)-\bm u_A(\bm R))
]^2
\right\}\,,
\end{eqnarray}
\end{widetext}
where $\bm a_1=\bm\delta_1-\bm\delta_3$ and  
$\bm a_2=\bm\delta_2-\bm\delta_3$, represent
the vectors defining the unit cell. The kinetic energy has the form
\begin{equation}
T=\sum_{\bm R}\frac {M_C}2[\dot{\bm u}_A(\bm R)]^2+
\frac{M_C}2[\dot{\bm u}_B(\bm R)]^2\,,
\end{equation}
where $M_C$ is the Carbon atom mass. The Lagrangian $L=T-V_1-V_2$ leads
to an eigenproblem of the form 
$M\omega^2\bm w=\bm D\bm w$, where
\begin{equation}
\bm w=
\left(
\begin{array}{c}
u_{Ax}\\
u_{Ay}\\
u_{Bx}\\
u_{By}
\end{array}
\right)\,,
\end{equation}
and $\bm D$ is the dynamical matrix reading
\begin{equation}
\bm D=
\left(
\begin{array}{cccc}
X_1 & F & A & C \\
F^\ast&X_2& C&B\\
A^\ast& C^\ast&X_1 & F\\
C^\ast&B^\ast&F&X_2
\end{array}
\right)\,,
\label{DM}
\end{equation}
with
\begin{eqnarray}
X_1 &=& \frac 3 2 \alpha +\frac {45}8\beta + \frac 98\beta\cos(\sqrt 3Q_ya)\,,
\\
X_2 &=&\frac 2 3\alpha +\frac {45}8\beta- \frac 3 8 \beta\cos(\sqrt 3Q_ya)
\nonumber\\
&+&\frac 3 2\beta \cos(3Q_xa/2)\cos(\sqrt 3 Q_ya/2)\,,\\
A&=&-\alpha e^{-iQ_xa}-\left(\frac {\alpha}2 +\frac{27}4\beta
\right)\cos(\sqrt 3Q_ya/2)e^{iQ_xa/2}\,,\\
B&=&
-\frac9 2\beta e^{-iQ_xa}-\left(
\frac 3 2 \alpha+\frac 9 4\beta
\right) \cos(\sqrt 3Q_ya/2)e^{iQ_xa/2}
\,,\\
C&=&-i\sqrt 3\left(\frac {\alpha}2-\frac 9 4\beta
\right)\sin(\sqrt 3Q_ya/2)e^{iQ_xa/2}\,,\\
F&=&i\beta\frac  {3\sqrt 3}8
[\sin(\sqrt 3Q_ya)-2\sin(\sqrt 3Q_y/2)e^{i3Q_xa/2}]\,,\\
\end{eqnarray}
where we have redefined $\beta$ as $\beta/a^2$ and $\bm Q=(Q_x,Q_y)$
is the momentum of the excitation. This model can be diagonalized 
numerically and its spectrum is represented in Fig. \ref{phonon}.

Although this model can easily be solved numerically, it is useful
to derive from it  a simple analytical model which helps in the
analytical calculation of the electron-phonon problem. To do so,
we follow Suzuura and Ando\cite{Suzuura} and introduce two effective models,
one for the acoustic and the other for the optical excitations, 
associated with the fields
$\bm u=(\bm u_A+\bm u_B)/\sqrt 2$ and $\bm v=(\bm u_A-\bm u_B)/\sqrt 2$,
respectively.   The effective Hamiltonian for the acoustic modes has
the form
\begin{equation}
H=
K_1a^2
\left(
\begin{array}{cc}
Q^2_x & Q_xQ_y\\
Q_xQ_y&Q^2_y
\end{array}
\right)
+K_2a^2
\left(
\begin{array}{cc}
Q^2_x +Q^2_y& 0\\
0&Q^2_x +Q^2_y
\end{array}
\right)\,,
\end{equation}
with $K_1=3\alpha/8$ and $K_2=27\alpha\beta/(8\alpha+36\beta/2)$.
The eigenmodes of this effective Hamiltonian are
\begin{equation}
M_C\omega^2=K_2(Qa)^2\,,
\label{EGMA1}
\end{equation}
and
\begin{equation}
M_C\omega^2=(K_2+K1)(Qa)^2\,,
\label{EGMA2}
\end{equation}
with polarization vectors
\begin{equation}
\bm\epsilon_{TA}=i(-Q_y,Q_x)/Q\,,
\end{equation}
and
\begin{equation}
\bm\epsilon_{LA}=i(Q_x,Q_y)/Q\,,
\end{equation}
respectively. The velocity of the modes is
given by
\begin{equation}
v_{TA}=a\sqrt{K_2/M_C}\,,
\end{equation}
and
\begin{equation}
v_{LA}=a\sqrt{(K_1+K_2)/M_C}\,.
\end{equation}

The eigenmodes (\ref{EGMA1}) and (\ref{EGMA2})
are represented in Fig. \ref{phonon} by dashed lines.
They cannot be distinguished from the effective model described in
 Appendix \ref{AA}. The effective Hamiltonian for the optical modes
is given by

\begin{eqnarray}
H&=&K_0
\left(
\begin{array}{cc}
1 & 0\\
0&1
\end{array}
\right)
-K_3a^2
\left(
\begin{array}{cc}
Q^2_x & Q_xQ_y\\
Q_xQ_y&Q^2_y
\end{array}
\right)\nonumber\\
&-&K_4a^2
\left(
\begin{array}{cc}
Q^2_x +Q^2_y& 0\\
0&Q^2_x +Q^2_y
\end{array}
\right)\,,
\end{eqnarray}
with $K_0=3(\alpha+9\beta/2)$, $K_3=(3\alpha/8-27\beta/8)$, and 
$K_4=27\beta(\alpha+9/4\beta)/(8\alpha+36\beta)$. The eigenmodes
are 
\begin{equation}
M_C\omega^2=K_0-K_4(Qa)^2\,,
\label{EGMO1}
\end{equation}
and
\begin{equation}
M_C\omega^2=K_0-(K_3+K_4)(Qa)^2\,,
\label{EGMO2}
\end{equation}
with polarization vectors given by
\begin{equation}
\bm\epsilon_{TO}=i(-Q_y,Q_x)/Q\,,
\end{equation}
and
\begin{equation}
\bm\epsilon_{LO}=i(Q_x,Q_y)/Q\,,
\end{equation}
respectively. The modes (\ref{EGMO1}) and (\ref{EGMO2}) are represented
in Fig. (\ref{phonon}) by dashed lines. The dashed-dotted lines of the same Figure are derived from an effective model, discussed in Appendix \ref{AA}.

%----------------------------------------------------------------------------%
% Section 
%----------------------------------------------------------------------------%
\subsection{Electron-phonon interaction} 
We now address the question of the electron-phonon interaction. This comes
about because the hopping $t$ depends on the absolute distance between 
neighboring carbon atoms. We therefore have
\begin{eqnarray}
&&t[a+(\bm u_A(\bm R)-\bm u_B(\bm R+\bm\delta))\cdot\bm\delta/a]
\simeq t[a]
\nonumber\\
&+&\frac 1 a \frac {\partial t[a]}{\partial a}
(\bm u_A(\bm R)-\bm u_B(\bm R+\bm\delta))\cdot\bm\delta\,.
\label{hopping}
\end{eqnarray}
Replacing (\ref{hopping}) in the Hamiltonian (\ref{TB}) and introducing
the Fourier representation
\begin{equation}
a_\sigma(\bm R)=\frac {1}{\sqrt N_c}\sum_{\bm k}e^{i\bm k\cdot\bm R}
a_{\sigma}(\bm k)\,,
\end{equation}
and
\begin{equation}
u_A(\bm R)=\frac {1}{\sqrt N_c}\sum_{\bm Q}e^{i\bm Q\cdot\bm R}
u_A(\bm Q)\,,
\end{equation}
with similar equations for $b_\sigma(\bm R)$ and $u_B(\bm R)$, the 
electron-phonon interaction has form
\begin{eqnarray}
H_{e-ph}&=&-\frac 1 a \frac {\partial t}{\partial a}
\frac {1}{\sqrt N_c}\sum_{\bm Q,\bm k}\sum_{\sigma,\bm\delta}
[\bm u_A(\bm Q)-\bm u_B(\bm Q)e^{i\bm Q\cdot\bm\delta}]\cdot\bm\delta
\nonumber\\
&\times&[e^{i\bm k\cdot\bm\delta}a^\dag_\sigma(\bm k+\bm Q)b_\sigma(\bm k)
+e^{-i\bm k\cdot\bm\delta}
b^\dag_\sigma(\bm k)a_\sigma(\bm k-\bm Q)]\,.
\end{eqnarray}
Since we are interested in the effect of the phonons with momentum
close to the $\Gamma$-point, the phase $e^{i\bm Q\cdot\bm\delta}$
is expanded as $e^{i\bm Q\cdot\bm\delta}\simeq 1-i\bm Q\cdot\bm\delta$.
Introducing the optical modes $\bm v$, the electron-phonon interaction
with the optical phonon modes has the form

\begin{eqnarray}
H_{e-ph}^{opt}&=&-\frac 1 a \frac {\partial t}{\partial a}
\frac {1}{\sqrt N_c}\sum_{\bm Q,\bm k}\sum_{\sigma,\nu,\bm\delta}
\sqrt{\frac {\hbar}{M_C\omega_\nu(\bm Q)}}
\bm\epsilon_\nu(\bm Q)\cdot\bm\delta
\nonumber\\
&\times&(B^\dag_{-\bm Q,\nu}+B_{\bm Q,\nu})
[e^{i\bm k\cdot\bm\delta}a^\dag_\sigma(\bm k+\bm Q)b_\sigma(\bm k)
\nonumber\\
&+&e^{-i\bm k\cdot\bm\delta}
b^\dag_\sigma(\bm k)a_\sigma(\bm k-\bm Q)]\,.
\label{Hop}
\end{eqnarray}
Introducing the acoustic modes $\bm u$, the electron-phonon interaction
with the acoustic phonon modes has the form

\begin{eqnarray}
H_{e-ph}^{ac}&=&\frac i a \frac {\partial t}{\partial a}
\frac {1}{\sqrt N_c}\sum_{\bm Q,\bm k}\sum_{\sigma,\nu,\bm\delta}
\sqrt{\frac {\hbar}{4M_C\omega_\nu(\bm Q)}}
(\bm\epsilon_\nu(\bm Q)\cdot\bm\delta)
\nonumber\\
&\times&(\bm Q\cdot\bm\delta)(B^\dag_{-\bm Q,\nu}+B_{\bm Q,\nu})
[e^{i\bm k\cdot\bm\delta}a^\dag_\sigma(\bm k+\bm Q)b_\sigma(\bm k)
\nonumber\\
&+&e^{-i\bm k\cdot\bm\delta}
b^\dag_\sigma(\bm k)a_\sigma(\bm k-\bm Q)]\,.
\label{Hac}
\end{eqnarray}
In both Eqs. (\ref{Hop}) and (\ref{Hac}) the 
$B_{\bm Q,\nu} (B^\dag_{\bm Q,\nu})$ operators are
destruction (creation) operators of phonons of momentum $\bm Q$
and polarization $\nu$.

%----------------------------------------------------------------------------%
% Section 
%----------------------------------------------------------------------------%
\section{Electronic Green's function} 

The electronic Green's function in the Dirac cone approximation has the form
\begin{equation}
\trG = \frac {
\left(
\begin{array}{cc}
 i\omega_n-\Sigma(i\omega_n)& -t\phi(\bm k)\\
-t\phi^\ast(\bm k) &i\omega_n-\Sigma(i\omega_n)
\end{array}
\right)
}
{[i\omega_n-\Sigma(i\omega_n)][i\omega_n-\Sigma(i\omega_n)]-t^2\vert \phi(\bm k)\vert^2}\,,
\label{dyson_solved}
\end{equation}
with $\phi(\bm k)=\sum_{\bm \delta}e^{i{\bm k}\cdot{\bm \delta}}$.
The electronic self-energy shall be given by
\begin{equation}
\Sigma(i\omega_n)=\Sigma^{imp}(\bm K,i\omega_n)+\Sigma^{pho}(\bm
K,i\omega_n)\,,
\end{equation}
where $\Sigma_{imp}(\bm K,i\omega_n)=\Sigma_{unit}(i\omega_n)+\Sigma_{Coul}(\bm K,i\omega_n)$ represents the contribution
due to mid-gap states (unitary scatterers) as well as long-range Coulomb scatterers. The self-energy $\Sigma_{pho}(\bm K,i\omega_n)=\Sigma_{opt}(\bm K,i\omega_n)+\Sigma_{ac}(\bm K,i\omega_n)$ represents the contributions due to optical and acoustic phonons. We note that the self-energies originating from the electron-phonon and Coulomb
interaction are evaluated at the Dirac momentum $\bm K=2\pi(1/3,\sqrt 3/9)/a$. In the following, we will discuss the Green's functions due to the the various contributions.

%-------------------------------------------------------------------
% SubSection 
%-------------------------------------------------------------------
\subsection{Green's function with mid-gap states (unitary scatterers)} 
\label{GC}

The physical origin of mid-gap states in the spectrum of graphene is
varied. Cracks, edges, vacancies\cite{Ursel} are all possible sources for mid-gap
states. From an analytical point of view, these types of impurities
(scatterers) are easily modeled by considering the effect of
vacancies. We stress, however, that this route is chosen due to its
analytic simplicity. 

The effect of mid-gap states on the
conductivity of graphene was first considered by Peres {\it et al.}
\cite{PeresPRB} for the case a half-filled system.  Considering the
effect of a local scattering potential of intensity $\epsilon_0$, the
Green's function has the form of Eq. (\ref{dyson_solved}) with the
retarded self-energy
\begin{equation}
\Sigma_{unit}^{ret}(\omega)=\frac {n_i\epsilon_0}{\hbar}
\frac {1-\epsilon_0 F(\omega)-i\epsilon_0\pi R(\omega)}
{[1-\epsilon_0 F(\omega)]^2+[\epsilon_0\pi R(\omega)]^2}\,,
\end{equation}
where the functions $F(\omega)$ and $R(\omega)$ are defined by
\begin{equation}
\frac {1}{\hbar N_c}\sum_{\bm k}G(\bm k,\omega+i0^+)=F(\omega)
-i\pi R(\omega)\,.
\end{equation}  
Mid-gap states are obtained by making the limit
$\epsilon_0\rightarrow\infty$, which resembles the unitary
limit. Clearly, $R(\omega)$ is the density of states per spin per unit
cell.  Let us write $\Sigma_{unit.}^{ret}(\omega)$ as a sum of real
and imaginary parts,
$\Sigma_{unit}^{ret}(\omega)=\Sigma'(\omega)+i\Sigma''(\omega)$ (note
that $\Sigma''>0$). The functions $F(\omega)$ and $R(\omega)$ are
determined self-consistently through the numerical solution of the
following set of equations:
\begin{eqnarray}
F(\omega)-i\pi R(\omega)=
\frac {1}{t^2\pi\sqrt 3}[
\hbar\Sigma''(\Upsilon-i\Psi/2)
\nonumber\\
+(\hbar\omega-\hbar\Sigma')(\Psi/2-i\Upsilon)
]\,,
\end{eqnarray}
with $\Upsilon$ and $\Psi$ given by
\begin{equation}
\Upsilon = \sum_{\alpha=\pm 1}-\arctan\frac {\hbar\Sigma'-\omega\hbar}
{\hbar\Sigma''}+\arctan\frac {\hbar\Sigma'-\omega\hbar+\alpha D}{\hbar\Sigma''}\,,
\end{equation}
and
\begin{equation}
\Psi=\sum_{\alpha=\pm 1}
\log\frac {(\hbar\Sigma'')^2+(\hbar\Sigma'-\hbar\omega)^2}
{(\hbar\Sigma'')^2+(\hbar\Sigma'-\hbar\omega+\alpha D)^2}\,.
\end{equation}

\begin{figure}[!ht]
\begin{center}
\includegraphics*[scale=0.3]{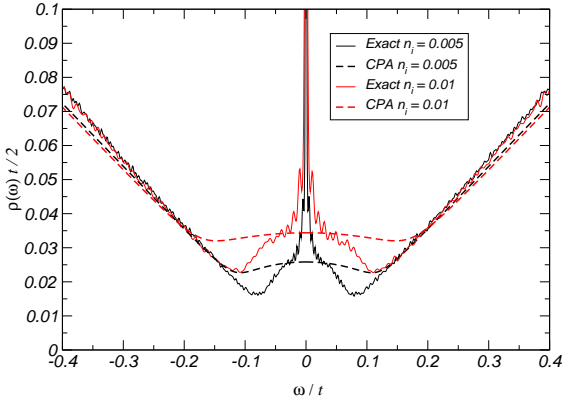}
\end{center}
\caption{
(color online) Density of states of graphene in the presence of mid-gap
states. The CPA calculation is compared with a numerical exact method.
The concentration of impurities is $n_i=0.005$ and $n_i=0.01$.
Here and in the following figures we use $t=3$ eV and a cutoff
energy of $D=7$ eV.
\label{dos}}
\end{figure}

In Figure \ref{dos}, we compare the density of states computed using
the coherent potential approximation (CPA) equations with that
obtained from a numerical exact method.  \cite{vitorprl} It is clear
that the CPA captures the formation of mid-gap states in a
quantitative way.  The main difference is the presence of a peak at
zero energy in the exact density of states, whose measure is
quantitatively negligible.

\begin{figure}[!ht]
\begin{center}
\includegraphics*[scale=0.3]{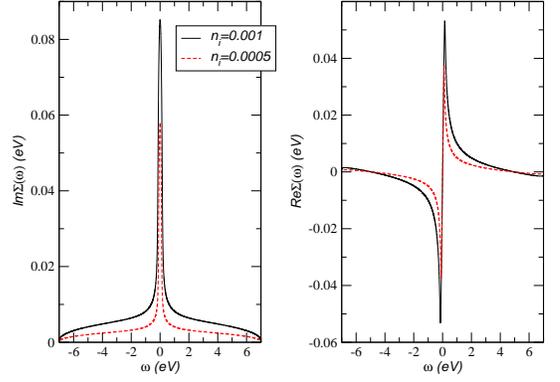}
\end{center}
\caption{
(color online) CPA calculation of the self-energy
$\hbar\Sigma_{unit}^{ret}(\omega)$ for two impurity concentrations, 
$n_i=0.001$ and $n_i=0.0005$. The left panel shows the imaginary
part and the right one shows the real part of the self-energy.
\label{cpase}}
\end{figure}

In Figure \ref{cpase} we depict the self-energy calculation using the CPA
equations, for two different values of the impurity concentration. It is clear
that $\Sigma''(\omega)$ increases close to zero energy leading to a broadening
of the electronic states close to the Dirac point.

\subsection{Green's function with Coulomb impurities} 
\label{GFC}

It has been argued that charged impurities are crucial to understand
the transport properties of graphene on top of a silicon oxide
substrate.\cite{Nomura,Adam07,Chen} In what follows, we compute the
electronic self-energy due to charge impurities, using second order
perturbation theory in the scattering potential. Electronic
scattering from an impurity of charge $Ze$ leads to a term in
the Hamiltonian of the form
\begin{equation}     
V=-\sum_{\bm R,\sigma}\frac {Ze^2}{\sqrt{d^2+\bm R^2}}
[a^\dag_\sigma(\bm R)a_\sigma(\bm R)+b^\dag_\sigma(\bm R)b_\sigma(\bm R)]\,.
\end{equation}
In momentum space $V$ reads
\begin{equation}
V=\frac 1 {N_c}\sum_{\bm p, \bm q,\sigma}V_0(\bm q)
[a^\dag_\sigma(\bm p)a_\sigma(\bm p+\bm q)+
b^\dag_\sigma(\bm p)b_\sigma(\bm p+\bm q)],
\end{equation}
where $V_0(\bm q)$ reads
\begin{equation}
V_0(\bm q) = -\sum_{\bm R}\frac {Ze^2e^{i\bm R\cdot\bm q}}
{\sqrt{d^2+\bm R^2}}\,.
\end{equation}

With $\trG^0(\bm k,i\omega_n)$ the bare and $\trG(\bm k,\bm p,
i\omega_n)$ the full Green's functions, the Dyson equation due to one
Coulomb impurity reads
\begin{eqnarray}
\trG(\bm k,\bm p,\omega_n)&=&\delta_{\bm k,\bm p}+\trG^0(\bm k,i\omega_n)\nonumber\\
&\times&\frac{1}{\hbar N_c}\sum_{\bm k'}V_0(\bm k-\bm k')\trG(\bm k',\bm p,\omega_n)\;.
\end{eqnarray}

If we consider a finite density per unit cell, $n^C_i$, and incoherent scattering between impurities, the second-order self-energy is given by
\begin{equation}
\Sigma_{Coul}^{ret}(\bm k,i\omega_n) =\frac {n^C_i} {\hbar^2 N_c}
\sum_{\bm p}V^2(\bm k-\bm p)\trG^0(\bm p,i\omega_n)\,,
\label{SEC}
\end{equation}
where a term of the form $n^C_iV(0)/\hbar$ was absorbed in the chemical 
potential, since it corresponds to an energy shift only. Note that
we have replaced $V_0(\bm q)$ by $V(\bm q)$, which corresponds to include
the effect of electronic screening in the calculation. The form of 
$V(\bm q)$ is (in S.I. units)\cite{StauberBZ}
\begin{equation}
V(\bm q) = -\frac{Z e^2}{2\epsilon_0\epsilon A_c}\frac{e^{-qd}}{q+\gamma}\,,
\end{equation} 
where $\epsilon=3.9$ is the Silicon Oxide relative permittivity, $d$
is the distance from the charge to the graphene plane, and $\gamma$ is
given by
\begin{equation}
\gamma = \frac {\rho(\mu)e^2}{2\epsilon_0\epsilon A_c}\,,
\end{equation}
where $\rho(\mu)$ is the self-consistent density of states as computed
from the CPA calculation ($A_c=3\sqrt{3}a^2/2$ is the area of the unit
cell). 

The self-energy (\ref{SEC}) is dependent both on the momentum
$\bm k$ and on the frequency. However, we are interested on the effect
of the self-energy for momentum close to the Dirac point.  Within this
approximation the imaginary part of the retarded self-energy becomes
diagonal and momentum independent, reading ($d\simeq 0$)
\begin{equation}
\hbar\Im\Sigma_{Coul}^{ret}(\bm K,\omega)\simeq -\frac {Z^2e^4}
{4A_c^2\epsilon_0^2\epsilon}
\frac {n^C_i}{\sqrt 3 t^2} \vert\hbar\omega\vert \left(
\frac {2\vert\hbar\omega\vert}{3ta}
+\gamma\right)^{-2}\,.
\end{equation}

The self-energy contribution due to Coulomb impurities is the most relevant one in order to fit the experimental data of Ref. [\onlinecite{Basov}] and is shown on the right hand side of Fig. \ref{S0fit}.

\subsection{Green's function with phonons}
Following the same procedure as in the previous subsection, the self-energy due to optical phonons within first order perturbation theory is given by
\begin{eqnarray}
\Sigma^{opt}(\bm K,i\omega_n)=
-\frac 9 2 \left(
\frac {\partial t}{\partial a}
\right)^2\frac 1 {\hbar M_C\omega_0}
\frac 1 {N_c}\sum_{\bm Q}
\nonumber\\
\times\frac 1{\beta\hbar}\sum_m
D^0(\bm Q,i\nu_m)G^0(\bm K-\bm Q,i\omega_n-i\nu_m)\,.
\end{eqnarray}

The analytical form of the self-energy due to acoustic phonons reads
\begin{eqnarray}
\Sigma^{ac}(\bm K,i\omega_n)=
-\frac 9 {16} \left(
\frac {\partial t}{\partial a}
\right)^2
\frac 1 {N_c}\sum_{\bm Q,\nu}
\frac {(Qa)^2}{4\hbar M_C\omega_\nu(\bm Q)}
\nonumber\\
\times\frac 1{\beta\hbar}\sum_m
D^0(\bm Q,i\nu_m)G^0(\bm K-\bm Q,i\omega_n-i\nu_m)\,.
\end{eqnarray}

The unperturbed Green's functions have the form
\begin{equation}
D^0(\bm Q,i\nu_m)=\frac {2\omega_\nu(\bm Q)}{(i\nu_m)^2-[\omega_\nu(\bm Q)]^2}
\,,
\end{equation}
and
\begin{equation}
 G^0(\bm k,i\omega_n) = \frac{i\omega_n}
{(i\omega_n)^2-t^2\vert \phi(\bm k)\vert^2}\,,
\label{G0}
\end{equation}
and $\omega_0=\sqrt{K_0/M_C}$.  The Matsubara summation over the
frequency $\nu_m$ is done using standard methods. 

\subsubsection{Effect of Disorder: Fermionic Propagator}
If we include the effect of disorder, the unperturbed Green's function
$G^0(\bm K-\bm Q,i\omega_n-i\nu_m)$ should be replaced by the dressed
Green's function due to the impurities. For the (relevant) case of
optical phonons, where the phonon dispersion is approximates by
$\omega_\nu(\bm Q)\simeq\omega_0$, the calculations are simple to do
and the result for the imaginary part of the self-energy due to
optical phonons is
\begin{eqnarray}
&&\hbar \Im \Sigma^R_{op}(\bm K,\omega)=
-\frac 9{2\sqrt 3 \pi}\left(\frac {\partial t}{\partial a}\right)^2
\frac {\hbar}{M_C\omega_0 t^2}\sum_{\alpha=\pm 1}\sum_{\beta=\pm 1}
\nonumber\\
&&\left(
2A_\alpha\arctan\frac {A_\alpha}{I_a^\alpha}
-2A_\alpha\arctan\frac {A_\alpha-s\beta D}{I_a^\alpha}
\right.\nonumber\\
&&+\left.
2I_a^\alpha\log\frac {(A_\alpha-s\beta D)^2+(I_a^\alpha)^2}
{A_\alpha^2+(I_a^\alpha)^2}
\right)\times\nonumber\\
&&\left[
n_B(\hbar\omega_0)+(1-\alpha)/2+
\alpha n_F(\hbar\omega+\alpha\hbar\omega_0-\mu)
\right]
\,,
\end{eqnarray}
with $s={\rm sign\,}A$,
\begin{eqnarray}
A_\alpha &=&\hbar\omega+\alpha\hbar\omega_0- 
\hbar\Sigma'(\omega+\alpha\omega_0)\,,
\\
I_a^\alpha&=&\hbar\Sigma''(\omega+\alpha\omega_0)\,,
\end{eqnarray}
and $n_B(x)$ and $n_F(x)$ the Bose and Fermi functions, respectively.

\begin{figure}[!ht]
\begin{center}
\includegraphics*[scale=0.3]{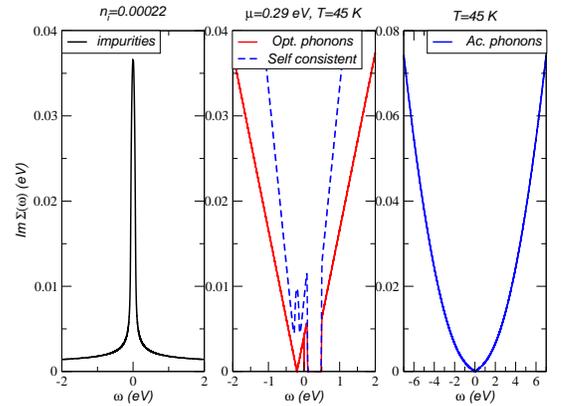}
\end{center}
\caption{
(color online) Electronic self-energy due to unitary scatterers and
acoustic and optical phonons. Only the imaginary part is represented.
The impurity concentration is $n_i=0.0002$. The self-energy due to optical
phonons depends on temperature and on the chemical potential $\mu$.
We have chosen $\mu=0.29$ eV and $T=45$ K. In the central panel
we show the calculation for self-energy due to optical phonons
both using the bare electronic Green's function (solid line) and
the Green's function with mid-gap states (dashed line). The
acoustic self-energy is independent of the chemical potential and is represented
for a temperature of $T=45$ K.
\label{SE}}
\end{figure}

The  self-energy due to optical phonons,
computed using the disordered electronic Green's
function, is compared with the same quantity computed using the 
bare electronic Green's function in the central panel of 
Fig. \ref{SE}. It is clear that the imaginary part of the self-energy
has a larger value when the disordered Green's function is used.
However, in  the region of frequencies  
$\mu-\hbar\omega_0<\hbar\omega<\hbar\omega_0+\mu$, at $T=0$, the imaginary
part coming from the optical phonons is zero, both when one uses
the bare and the disordered Green's functions. This is due to the
arguments of the Bose and Fermi functions.

In Fig. \ref{SE}, we depicted the self energy of the short ranged impurities 
together with those due to acoustic and optical phonons. It clear that
the effect of acoustic phonons is negligible at low energies.
The self-energy due to optical phonons depends on the chemical potential,
and is represented for a gate voltage of $V_g=71$ V ($\mu=0.29$ eV).

\subsubsection{Effect of Disorder: Phononic Propagator}
To be consistent, also the phonon propagator has to be dressed due to its interaction with impurities. The phonon propagator shall be renormalized within the
RPA-approximation, i.e.,
\begin{align}
D_\nu^{RPA}(i\omega_n)=\frac{2\omega_\nu}{(i\omega_n)^2-(\omega_\nu)^2-2\omega_\nu\Pi_\nu(i\omega_n)}
\end{align}
where the first order of the phononic self-energy $\Pi_\nu(\omega_n)$ is proportional to the polarization defined as
\begin{align}
P^{(1)}(\omega_n)=\lim_{\vec q\rightarrow0}\frac{1}{A_s}\int_0^{\hbar\beta}d\tau e^{i\omega_n\tau}\langle T_\tau \rho(\vec q,\tau)\rho(-\vec q,0)\rangle\;,
\end{align}
with $\rho(-\vec q)$ denoting the density operator. Explicitly, we get for the imaginary part of the retarded phononic self-energy
\begin{align}
\Im\Pi_\nu^{ret}(\omega)&=\frac{\hbar}{M\omega_\nu}\left(\frac{\partial t}{\partial a}\right)^2\frac{18}{\sqrt{3}\pi t^2}\int\frac{d\omega_1}{2\pi}\Theta(\omega_1,\omega)\nonumber\\
&\times\left[n_F(\omega_1\hbar-\mu)-n_F(\omega_1\hbar+\omega\hbar-\mu)\right]\;,
\end{align}
where the dimensional function $\Theta(\omega_1,\omega)$ is given in Appendix \ref{AB}.

In Fig. \ref{Disorder}, the self-energy due to electron-phonon scattering (left) and the conductivity (right) are shown as they result using the bare (dashed) and dressed (full) phonon propagator. Since the effect is hardly appreciable, we show also the results where the phononic self-energy has been multiplied by a factor 10 (dotted-dashed). The renormalization of the phonon propagator due to disorder is thus negligible and the results of the following section will be obtained using the bare phonon propagator.

\begin{figure}[!ht]
\begin{center}
\includegraphics*[scale=0.3,angle=-90]{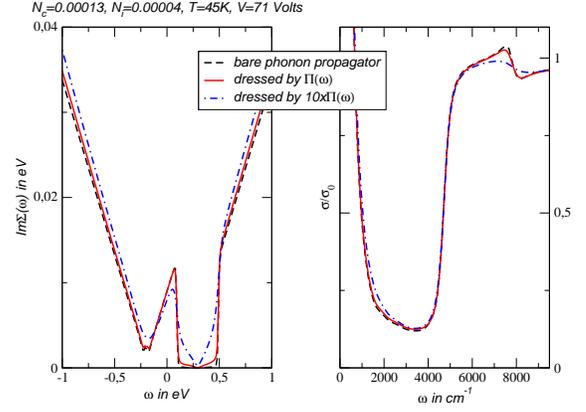}
\end{center}
\caption{
(color online) Electronic self-energy due to electron-optical phonon interaction $\Im\Sigma_{opt}^{ret}$ (left) and optical conductivity $\sigma$ (right) for various phonon propagators. The electronic propagator is dressed by the self-energies due to Coulomb ($n_C=0.00013$) and impurity scattering ($n_i=0.00004$). The temperature is $T=45$K and the applied gate voltage $V=71$Volts. 
\label{Disorder}}
\end{figure}

%----------------------------------------------------------------------------%
% Section 
%---------------------------------------------------------------------------%
\section{The DC and AC conductivity } 
\label{OC}
In this section, we discuss the transport properties of graphene due to the various sources of one-particle scattering. This is done within the Kubo formalism.
\subsection{The Kubo formula}

The Kubo formula for the conductivity is given by 
\begin{equation}
\sigma_{xx}(\omega) = \frac {< j^D_x>}{iA_s(\omega + i0^+)}+
\frac {\Lambda_{xx}(\omega + i0^+)}{i\hbar A_s(\omega + i0^+)}\,,
\end{equation}
with $A_s=N_cA_c$ the area of the sample, and $A_c=3\sqrt 3 a^2/2$ 
the area of the unit cell,
from which it follows that
\begin{equation}
\Re\sigma_{xx}
(\omega) = D\delta(\omega) + \frac {\Im \Lambda_{xx}(\omega + i0^+)}
{\hbar\omega A_s}\,,
\end{equation}
and
\begin{equation}
\Im\sigma_{xx}
(\omega) = -\frac {< j^D_x>}{A_s\omega} - \frac {\Re \Lambda_{xx}(\omega + i0^+)}
{\hbar\omega A_s}\,,
\end{equation}

where $D$ is the charge stiffness which reads
\begin{equation}
D= -\pi \frac {<j^D_x>}{A_s} -\pi\frac {\Re \Lambda_{xx}(\omega + i0^+) }
{\hbar A_s}\,.
\label{DW}
\end{equation}
The function $\Lambda_{xx}(\omega + i0^+)$ is obtained from the 
Matsubara current-current correlation function, defined as
\begin{equation}
\Lambda_{xx}(i\omega_n) = \int_0^{\beta}d\,\tau e^{i\omega_n\tau}
<T_{\tau} j^P_{x}(\tau)j^P_x(0)>\,.
\end{equation}
The calculation of the conductivity amounts to the determination of the
current-current correlation function.

%----------------------------------------------------------------------------%
% Section
%----------------------------------------------------------------------------%
\subsection{The real part of the DC conductivity}
The real part of the DC conductivity is given by
\begin{equation}
\Re\sigma(\mu)=-\frac {2e^2}{\pi h}\int {d\epsilon}
K(\epsilon)\frac{\partial f(\epsilon-\mu)}{\partial \epsilon}\,,
\label{SW_DC}
\end{equation}
where $f(x)$ is the Fermi function and $K(\epsilon)$ is a
dimensionless function that depends on the full self-energy.
In the limit of zero temperature the derivative of the Fermi function tends to a delta-function
and the conductivity is given by
\begin{equation}
\Re\sigma(\mu)=\frac {2e^2}{\pi h}K(\mu)\,,
\label{SWmu}
\end{equation}
 with $K(\mu)$ given by
\begin{widetext}
\begin{eqnarray}
K(\mu)&=&\frac{1}{2I_a}\sum_{s=\pm 1}\Big(
\frac{D(sA+D)I_a}{(A+sD)^2+I_a^2}+ A\arctan\frac{A}{I_a}-A\arctan\frac{A+sD}{I_a}
\Big) \nonumber\\
&+&\frac{I_a}{2A}\Big(
2\arctan\frac {A}{I_a}-\arctan{A-D}{I_a}\nonumber
-\arctan\frac{A+D}{I_a}
\Big)\,,\nonumber
\end{eqnarray}
\end{widetext}
and 
\begin{equation}
 I_a=\hbar\Im\Sigma(\mu)\,,\,
 A=\mu-\hbar\Re\Sigma(\mu)\,
\end{equation}
and $D$ the cutoff bandwidth. 

\begin{figure}[!ht]
\begin{center}
\includegraphics*[scale=0.3]{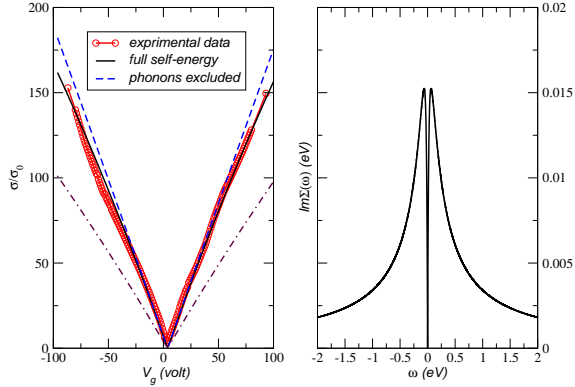}
\end{center}
\caption{(color online) Left panel: Experimental data (circles) from
Ref. [\onlinecite{Basov}], conductivity with the full self-energy
(solid line) due to phonons, mid-gap states, and charged impurities,
conductivity (dashed-line) with self-energy due to mid-gap states and
charged impurities, only.  The parameters are $T=45 $ K, $n_i=4\times
10^{-5}$, $n_i^C=7.5\times 10^{-5}$, and $d=0$.  The dashed-dotted
line is the same as the solid line but with $n_i^C=1.3\times 10^{-4}$.
Right panel: Imaginary part of the self-energy, $\hbar
\Im\Sigma_{Coul}^{ret}(\bm K,\omega)$, due to charged impurities, the most dominant contribution.
\label{S0fit}}
\end{figure}

In Fig. \ref{S0fit} we plot $\sigma_{DC}(\mu)$ as function of the gate
voltage, $V_g$, considering the effects of both charged impurities, mid-gap
states and acoustic phonons (the optical phonons do not contribute to
$\sigma_{DC}(\mu)$). It is clear that the linear behavior of  
$\sigma_{DC}(\mu)$ as function of $V_g$ is recovered. The theoretical curves
are plotted together with the experimental data of Ref. [\onlinecite{Basov}]. After having fitted the experimental curves, it is clear that the most dominant contribution to the DC conductivity is coming from Coulomb impurities.
 
\begin{figure}[!ht]
\begin{center}
\includegraphics*[scale=0.3]{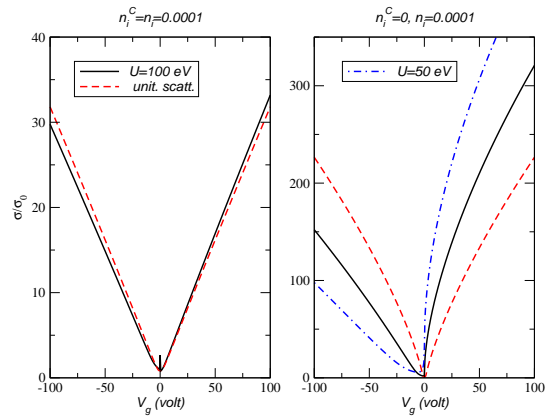}
\end{center}
\caption{ (color online) Left: Conductivity $\sigma(\mu)$, in units of $\sigma_0=\pi e^2/(2h)$,
considering both short range and charged scatterers. Righ:  Conductivity $\sigma(\mu)$ when
the influence  of charge scatterers is removed.
\label{Fig_s0}}
\end{figure}

Let us now further discuss the conductivity as function of the gate
voltage $V_g$, which relates to the chemical potential as $V_g\propto
\mu^2$.  In Figure \ref{Fig_s0} we show $\sigma(\mu)$ as function of
$V_g$ considering both charged impurities and short range scatterers
(left panel), having the same impurity concentration. We note that the
conductivity, albeit mostly controlled by charged impurities still has
finger prints of the finite $\epsilon_0$ scatterers, due to the asymmetry
between the hole (negative $V_g$) and particle (positive $V_g$)
branches. The conductivity follows closely the relation
$\sigma(\mu)\propto V_g$, except close to the Dirac point where is
value its controlled by the short range scattering.

If we suppress the scattering due to charged impurities, which should
be the case in suspended graphene, only the scattering due to short
range scatterers survive.  In this case the right panel of
Fig. \ref{Fig_s0} shows that there is a strong asymmetry between the
hole and the particle branches of the conductivity curve even for a
value of $\epsilon_0$ as large as 100 eV.  Moreover, the smaller the
value of $\epsilon_0$ the larger is the asymmetry in the conductivity
curve. We note that asymmetric conductivity curves were recently
observed. \cite{Andrei}  On the contrary, if the experimental data
shows particle-hole  symmetry of $\sigma(\mu)$ around the Dirac
point, then the dominant source of scattering is coming from very
strong short range potentials, i.e., scatterers that are in the
unitary limit.

\subsection{The real part of the AC conductivity}
The finite frequency part of the conductivity is given by
\begin{equation}
\Re\sigma(\omega)=\frac {2e^2}{\pi h}\int\frac {d\omega_1}
{\omega}\Theta(\omega_1,\omega)[f(\hbar\omega_1-\mu)-
f(\hbar\omega_1+\hbar\omega-\mu)]\,,
\label{SW}
\end{equation}
where $f(x)$ is the Fermi function and $\Theta(\omega_1,\omega)$
is a dimensionless function given in  Appendix \ref{AB}. As can be
see from this Appendix, the function $\Theta(\omega_1,\omega)$
depends on the self energy, which is due both to impurities and phonons.
%-------------------------------------------------------------------------------------------------------------------
% Optical conductivity without Coulomb scatterers
%-------------------------------------------------------------------------------------------------------------------
\subsubsection{Optical conductivity without Coulomb scatterers}
\begin{figure}[!ht]
\begin{center}
\includegraphics*[scale=0.3]{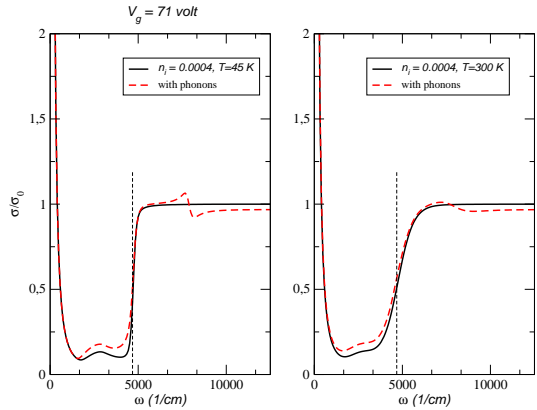}
\end{center}
\caption{ (color online) Optical conductivity of graphene at two
different temperatures $T=$45, 300 K. Each panel has two curves, the
solid curve is the conductivity with only short-ranged impurities, the
dashed curve is the conductivity with both short-ranged impurities and
phonons, with the self-energy due to phonons computed with the full
Green's function. The chemical potential is that associated with a
gate voltage of $V_g=71$ V. The vertical, dashed line marks twice this
value.
\label{somega}}
\end{figure}

We will first discuss the optical conductivity without Coulomb
scatterers since they should not be present in suspended graphene.  In
Figure \ref{somega}, we plot the conductivity of a graphene plane in
units of $\sigma_0=\frac{\pi}{2} e^2/h$. The calculation is made at two
different temperatures, $T=45$ K and $T=300$ K, and for a density of
impurities $n_i=0.0004$. The calculation compares the conductivity
with and without the effect of the phonons. The main effect induced by
short-ranged impurities is the existence of a finite light-absorption in the
frequency range $0<\hbar\omega<2\mu$. The optical phonons increase the
absorption in this frequency range. We have checked that the effect of
the acoustic phonons is negligible.  The optical phonons also induce a
conductivity larger than $\sigma_0$ for frequencies above
$\omega=2\mu$. The effect is more pronounced at low temperatures.

\begin{figure}[!ht]
\begin{center}
\includegraphics*[scale=0.3]{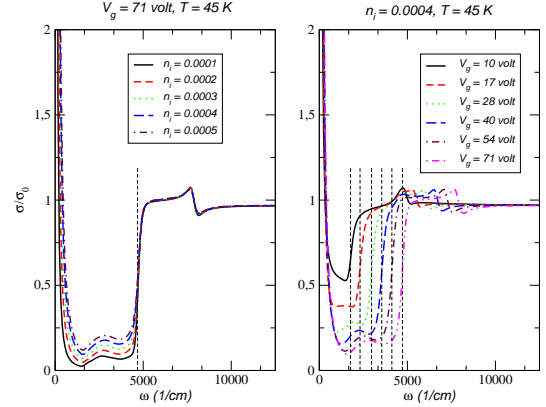}
\end{center}
\caption{ (color online) Optical conductivity of graphene at
temperature $T=$45 K. The left hand side shows curves for various
short-ranged impurity densities $n_i$. The right hand side shows
curves for various gate voltage $V_g$.  The vertical, dashed lines
mark twice the chemical potential.
\label{somega_II}}
\end{figure}

In Figure \ref{somega_II}, we again plot the conductivity of a
graphene plane in units of $\sigma_0=\frac{\pi}{2} e^2/h$ and at
temperature $T=45$ K. This time we compare different impurity
densities $n_i$ (left hand side) and gate voltages/chemical potentials
(right hand side). There is more absorption for frequencies in the
region $0<\hbar\omega<2\mu$ the larger the impurity concentration
is. This is because the number of mid-gap states is proportional to
$\sqrt n_i$.\cite{PeresPRB} For larger gate-voltage a plateau is
reached at frequencies in $0<\hbar\omega<2\mu$. For small gate
voltages, $V_g$, the absorption in the region $0<\hbar\omega<2\mu$ is
larger than for larger $V_g$.

\begin{figure}[!ht]
\begin{center}
\includegraphics*[scale=0.3]{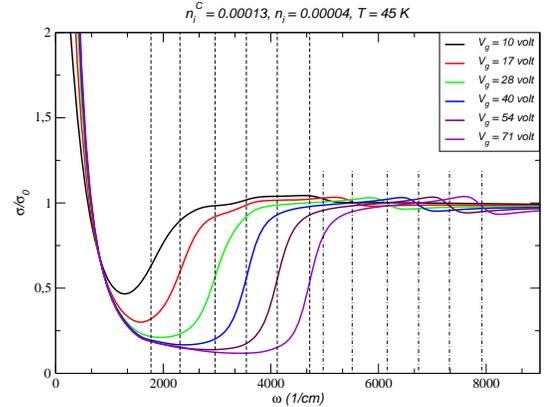}
\end{center}
\caption{(color online)  Real part of the optical
conductivity including the effect of phonons, midgap states and
charged impurities. The parameters
are $T=45 $ K, $n_i=4.0\times 10^{-5}$, and $n_i^C=1.3\times 10^{-4}$. The dashed vertical lines correspond to $\hbar\omega=2\mu$ and the shorter dotted-dashed ones to $\hbar\omega=2\hbar\omega_0+2\mu$.
\label{Fig_SW}}
\end{figure}

\begin{figure}[!ht]
\begin{center}
\includegraphics*[scale=0.3]{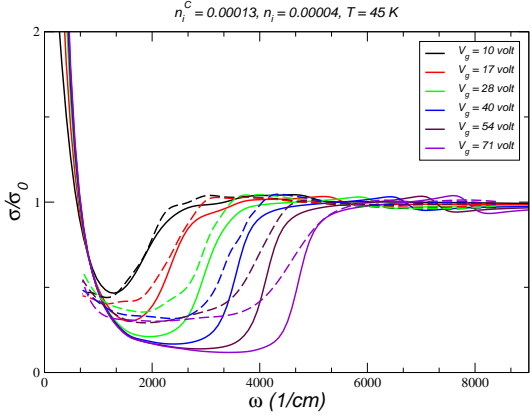}
\end{center}
\caption{(color online) Real part of the optical conductivity
including the effect of phonons, mid-gap states and charged
impurities, compared with the experimental data from
Ref. [\onlinecite{Basov}]. The parameters are $T=45 $ K,
$n_i=4.0\times 10^{-5}$, $n_i^C=1.3\times 10^{-4}$, and $d=0$.
\label{SWfit}}
\end{figure}
%-------------------------------------------------------------------------------------------------------------------
% Optical conductivity with Coulomb scatterers
%-------------------------------------------------------------------------------------------------------------------
\subsubsection{Optical conductivity with Coulomb scatterers}
We will now discuss the optical conductivity with Coulomb scatterers which are generally present in graphene on a substrate \cite{Chen}.
In Figure \ref{Fig_SW}, we plot the conductivity of a graphene plane in
units of $\sigma_0=\frac{\pi}{2} e^2/h$, including the effect of midgap
states, charged impurities and phonons. Again, the main feature is that the
conductivity is finite in the range $0<\hbar\omega<2\mu$ and increases
as the gate voltage decreases. We choose the concentration of midgap
states in Fig. \ref{Fig_SW} to be one order of magnitude smaller than
the one of Coulomb scatterers\cite{StauberBZ}, and therefore the
conductivity is mainly controlled by phonons and charged
impurities.

Another feature of the curves in Fig. \ref{Fig_SW} is the large
broadening of the interband transition edge at $\hbar\omega=2\mu$
(indicated by vertical dashed lines). Note that this broadening is not
due to temperature but to charged impurities, instead.  In fact, the
broadening for all values of $V_g$ is larger when the conductivity is
controlled by charged impurities rather than by midgap states.

The coupling to phonons produces a feature centered at
$2\hbar\omega_0+2\mu$ where $\hbar\omega_0$ denotes the LO-phonon
energy corresponding to the wave number 1600cm$^{-1}$ (indicated by
shorter dotted-dashed vertical lines). For gate voltages with
$\mu<\hbar\omega_0$ there appears a similar feature at
$2\hbar\omega_0-2\mu$ which is not washed out by disorder. The optical
phonons thus induce a conductivity larger than $\sigma_0$ around these
frequencies. This effect is washed out at larger temperatures and
$\sigma<\sigma_0$.

All these effects are consistent with the recent infrared
measurements of graphene on a SiO$_2$ substrate \cite{Basov} which are
shown as dashed lines in Fig. \ref{SWfit}. Notice that there is only
one fitting parameter involved which is adjusted by the conductivity
curve at zero chemical potential. Whereas for low gate
voltage the agreement is good, there is considerable weight missing
for higher gate voltage ($V>20$ Volts). Nevertheless, all theoretical
lines predict lower conductivity than the experimental measurements,
such that our model with only one
fitting parameter is consistent. Since we have included in our
calculation all possible one-particle scattering mechanisms, the
missing weight at large gate voltages could be attributed to
electron-electron interactions in graphene, which become important at
this electronic densities.
%-------------------------------------------------------------------------------------------------------------------
% Discussion and conclusions
%-------------------------------------------------------------------------------------------------------------------
\section{Discussion and conclusions}

In this paper we have computed the optical conductivity of graphene at
finite chemical potential, generalizing the results of Ref.
[\onlinecite{PeresPRB}]. The calculation includes both the effect of
disorder (mid-gap states and charged impurities, which have a
different signature in the DC conductivity\cite{Schliemann}) and the
effect of phonons (optical and acoustic). It is shown that at low
temperatures the effect of acoustic phonons in negligible, since it
induces an imaginary part of the electrons' self-energy that is much
smaller than the imaginary part induced by the impurities. For a
discussion based on the Boltzmann equation, see
Ref. \onlinecite{StauberBZ}.

The imaginary part induced by optical phonons is of the order of the
imaginary part induced by impurities. Still, optical phonons are only
important in the calculation of the real part of $\sigma(\omega)$,
they play no role in the calculation of $\sigma_{DC}(\mu)$ in the
temperature range $T\in[0,300]$ K. The self-energies due to mid-gap
states and due to optical phonons are in a sense complementary, since
the imaginary part coming from impurities is large at the Dirac point,
whereas the imaginary part coming from the optical phonons increase
linearly with the energy away from the Dirac point. The imaginary part
of the self energy due to charged impurities is a non-monotonous
function of the energy, growing first linearly but changing to a
decaying behavior of the form $1/\omega$, for large energies.  

In Section \ref{OC}, we have discussed the different scattering
mechanisms separately, since for suspended graphene Coulomb scatterers
will be absent. Since vacancies, corresponding to an infinite
potential where particle-hole symmetry is restored, are unlikely, we
model the short-ranged potentials due to cracks, ripples, etc. by
large, but finite short ranged potential. For the DC conductivity,
this leads to an asymmetry of the hole- and electron-doped regime.

The most notable effect on the optical conductivity of both the
optical phonons and the impurities is the induction of a finite energy
absorption in the energy range $0<\hbar\omega<2\mu$, a region where
the clean theory predicts a negligible absorption.  The optical
phonons also induce a conductivity larger than $\sigma_0$ around
$\hbar\omega\simeq 2\mu$. It is interesting to note that for
frequencies away from the Dirac point the imaginary part of the
self-energy due to optical phonons is linear in frequency, a behavior
similar to that due to electron-electron interactions in graphene.

It is  clear from Fig. \ref{SWfit} that in general the
 calculated absorption in the range $0<\hbar\omega<2\mu$ is not as large as
the experimental one. It is also noticeable that for small gate voltages
there is a reasonable fit of both the absorption and of the broadening
of the of the step around $2\mu$. For large values  of the gate
voltage the calculated absorption is smaller than the measured one.
This suggests that some additional scattering mechanism is missing in the
calculation of the optical conductivity. The missing mechanism has
to be more effective at large gate voltage. A possibility are plasmons
of the type found in Ref. [\onlinecite{Mikhailov}]. Another possibility is that water molecules which are especially active in the infrared regime, are contributing to the missing weight.  

In synthesis we have provided a complete and self-consistent
description of the optical conductivity of graphene on a substrate
(including Coulomb scatterers) and suspended (without Coulomb
scatterers). Our results are in qualitative agreement with the
experimental results. To meet a quantitative agreement further
research (also on suspended graphene) is necessary, but water
molecules underneath the graphene sheet are likely to account for the
missing weight.

\section*{Acknowledgments}
We thank D. N. Basov, A. K. Geim, F. Guinea, P. Kim, and Z. Q. Li for
many illuminating discussions. We thank D. N. Basov, and Z. Q. Li for
showing their data prior to publication. We also thank Vitor Pereira for
the exact curves we give in Fig.  \ref{dos}. This work was supported by
the ESF Science Program INSTANS 2005-2010, and by FCT under the grant
PTDC/FIS/64404/2006.
%----------------------------------------------------------------------------%
% Apendices
%----------------------------------------------------------------------------%
\appendix
%----------------------------------------------------------------------------%
% Section
%----------------------------------------------------------------------------%
\section{An approximate phonon model}
\label{AA}
We start by expanding the matrix elements of the dynamical matrix (\ref{DM})
up to second order in the momentum $\bm Q$. We then introduce acoustic,
$\bm u=(\bm u_A+\bm u_B)/\sqrt{2}$, and optical, 
$\bm v=(\bm u_A-\bm u_B)/\sqrt{2}$, modes.  This procedure leads to the
following eigenvalue problem
\begin{equation} 
M_C\omega^2
\left(
\begin{array}{c}
\bm u\\
\bm v
\end{array}
\right)
=
\left(
\begin{array}{cc}
B_1&B_2\\
B_2^\dag&B_3
\end{array}
\right)
\left(
\begin{array}{c}
\bm u\\
\bm v
\end{array}
\right)\,,
\end{equation}
with
\begin{equation}
B_1 = \frac {3a^2}{32}
\left(
\begin{array}{cc}
y_1& 4\alpha Q_xQ_y\\
4\alpha Q_xQ_y& y_2
\end{array}
\right)\,,
\end{equation}
with $y_1=3(2\alpha+3\beta)Q_x^2+(2\alpha+9\beta)Q_y^2$ and 
$y_2=(2\alpha+9\beta)Q^2_x+3(2\alpha+3\beta)Q^2_y$. Further, we have

\begin{equation}
B_2 = \frac {3ia^2}8(2\alpha-9\beta)
\left(
\begin{array}{cc}
Q_x& Q_y\\
Q_y& Q_x
\end{array}
\right)\,,
\end{equation}
and
\begin{equation}
B_3 = \frac {3a^2}{32}
\left(
\begin{array}{cc}
y_3& (36\beta-4\alpha)Q_xQ_y \\
(36\beta-4\alpha)Q_xQ_y   & y_4
\end{array}
\right)\,,
\end{equation}
with 
$y_3=32\alpha+144\beta-(6\alpha+9\beta)Q^2_x-(2\alpha+45\beta) Q_y^2$
and 
$y_3=32\alpha+144\beta-(6\alpha+9\beta)Q^2_y-(2\alpha+45\beta) Q_x^2$\,.
The eigenvalue problem can be put in the form
\begin{eqnarray}
B_1\bm u+B_2\bm v=M_C\omega^2\bm u\\
B_2^\dag\bm u+B_3\bm v=M_C\omega^2\bm v
\end{eqnarray}
Let us first look at the acoustic modes. They corresponds to
$\omega\simeq 0$. In this case we can write
\begin{equation}
\bm v \simeq -\frac {B_2^\dag}{3(\alpha+9\beta/2)}\bm u\,,
\end{equation}
from which an eigenvalue equation for $\bm u$ follows:
\begin{equation}
B_1\bm u -\frac {B_2B_2^\dag}{3(\alpha+9\beta/2)}\bm u=M\omega^2\bm u\,.
\label{u}
\end{equation}
In the case of the optical modes one has
$M\omega^2\simeq 3\alpha+27\beta/2 $, from which we can write
\begin{equation}
\bm u\simeq \frac {B_2}{3(\alpha+9\beta/2)}\bm v\,,
\end{equation}
which leads to 
\begin{equation}
B_3\bm v +\frac {B_2^\dag B_2}{3(\alpha+9\beta/2)}\bm v=M\omega^2\bm v\,.
\label{v}
\end{equation}
The dashed-dotted lines in Fig. \ref{phonon} are the eigenvalues of
Eqs. (\ref{u}) and (\ref{v}).

%----------------------------------------------------------------------------%
% Section
%----------------------------------------------------------------------------%
\section{The function $\Theta(\omega_1,\omega)$}
\label{AB}
In Eq. (\ref{SW}) the dimensionless $\Theta(\omega_1,\omega)$
function was introduced. Let us define
\begin{eqnarray}
A&=&\hbar\omega_1-\hbar\Sigma'(\omega_1)\,,
\\
B&=&\hbar\omega_1+\hbar\omega-\hbar\Sigma'(\omega_1+\omega)\,,
\\
I_a&=&\hbar\Sigma''(\omega_1)\,,
\\
I_b&=&\hbar\Sigma''(\omega_1+\omega)\,,
\\
D_1&=&2[(A-B)^2+(I_a-I_b)^2]\times
\nonumber\\
&&[(A-B)^2+(I_a+I_b)^2]\,,
\\
D_2&=&2[(A+B)^2+(I_a-I_b)^2]\times
\nonumber\\
&&[(A+B)^2+(I_a+I_b)^2]\,.
\end{eqnarray}
The function  $\Theta(\omega_1,\omega)$ is expressed in terms of the above
auxiliary functions as follows
\begin{widetext}
\begin{eqnarray}
\Theta(\omega_1,\omega)&=&\frac 2 {D_1}
I_b(A^3-2A^2B-2BI_a^2+A(B^2+I_b^2+I_a^2))
\sum_{\alpha=\pm 1}\left(
\arctan\frac {A}{I_a}-\arctan\frac {A+\alpha D}{I_a}
\right)\nonumber\\
&+&
\frac 2 {D_1}
I_a(B^3-2B^2A-2AI_b^2+B(A^2+I_a^2+I_b^2))
\sum_{\alpha=\pm 1}\left(
\arctan\frac {A}{I_b}-\arctan\frac {A+\alpha D}{I_b}
\right)\nonumber\\
&+&\frac 2 {D_2}
I_b(A^3+2A^2B+2BI_a^2+A(B^2+I_b^2+I_a^2))
\sum_{\alpha=\pm 1}\left(
\arctan\frac {A}{I_a}-\arctan\frac {A+\alpha D}{I_a}
\right)\nonumber\\
&+&
\frac 2 {D_2}
I_a(B^3+2B^2A+2AI_b^2+B(A^2+I_a^2+I_b^2))
\sum_{\alpha=\pm 1}\left(
\arctan\frac {A}{I_b}-\arctan\frac {A+\alpha D}{I_b}
\right)\nonumber\\
&+&
\left(
\frac 1 {D_1}+\frac 1 {D_2}
\right)
I_aI_b(A^2-B^2+I^2_a-I^2_b)
\sum_{\alpha=\pm 1}\left(
\log\frac{A^2+I_a^2}{B^2+I_b^2}-\log\frac{(A-\alpha D)^2+I_a^2}
{(B-\alpha D)^2+I_b^2}
\right)\,.
\end{eqnarray} 
\end{widetext}

%----------------------------------------------------------------------------%
% References
%----------------------------------------------------------------------------%

%%%%%%%%%%%%%%%%%%%%%%%%%%%%%%%%%%%%%%%%%%%%%%%%%%%%%%%%%%%%%%%%%%%%%%%%%%%%%%
\end{document}